\documentclass[aps,prd,nofootinbib,reprint,superscriptaddress,longbibliography]{revtex4-1}
\usepackage{etoolbox}
\usepackage{dcolumn}
\usepackage{bm}
\usepackage{graphics}
\usepackage{xcolor}
\usepackage{float}
\usepackage{dirtytalk}
\usepackage{graphicx} 
\usepackage{blindtext}
\usepackage{amsmath}
\usepackage{amsfonts}
\usepackage{enumitem}
\usepackage{amssymb}
\usepackage{mathtools,halloweenmath}
\usepackage[%
colorlinks=true,
urlcolor=blue,
linkcolor=red,
citecolor=blue
]{hyperref}
\usepackage{physics}
\usepackage{multirow}
\usepackage{booktabs}
\usepackage{siunitx}
\NewDocumentCommand{\tens}{t_}
{%
	\IfBooleanTF{#1}
	{\tensop}
	{\otimes}%
}
\NewDocumentCommand{\tensop}{m}
{%
	\mathbin{\mathop{\otimes}\displaylimits_{#1}}%
}
\usepackage{natbib}
\begin{document}
\title{\bf A holographic realization of correlation and mutual information}
\vskip 1cm
\author{Ashis Saha}
\email{ashis.saha@bose.res.in}
 \author{Anirban Roy Chowdhury}
 \email{iamanirban@bose.res.in}
 \author{Sunandan Gangopadhyay}
 \email{sunandan.gangopadhyay@bose.res.in}
\affiliation{Department of Astrophysics and High Energy Physics,\linebreak
	S.N.~Bose National Centre for Basic Sciences,\linebreak
	 JD Block, Sector-III, Salt Lake, Kolkata 700106, India}	

	\begin{abstract}
		\noindent The status of the inequality existing between mutual information and (normalized) thermal two-point connected correlation function, namely, $I(A:B)\ge\frac{(\expval{\mathcal{O}_{A}\mathcal{O}_{B}}_{\beta}-\expval{\mathcal{O}_{A}}_{\beta}\expval{\mathcal{O}_{B}}_{\beta})^2}{2\expval{\mathcal{O}_{A}^2}_{\beta}\expval{\mathcal{O}_{B}^2}_{\beta}}$ has been explicitly probed by using the gauge/gravity correspondence. In the holographic analysis, the geodesic approximation for heavy operators ($\Delta\sim mR$) has been used. We observe that the study leads to some non-trivial insights depending upon the method of calculating the thermal object $\expval{\mathcal{O}^2}_{\beta}$. For a particular computed result of $\expval{\mathcal{O}^2}_{\beta}$ we propose that all of the existing quantum mechanical dependencies (correlations) and classical correlations between the subsystems $A$ and $B$ vanishes at two different separation lengths, namely, $sT|c$ and $sT|_I$ where $sT|_I>sT|_c$.
	\end{abstract}
\maketitle
\section{Introduction}
\noindent In classical information theory, mutual information (MI) is a measure that quantifies the amount of information shared between two random variables. It tells us how much knowing one variable reduces uncertainty about the other. It can be expressed in terms of entropy as
\begin{eqnarray}\label{CMI}
I(X:Y)=H(X)+H(Y)-H(X,Y)
\end{eqnarray}
where $H(X),~H(Y)$ are the Shannon (differential) entropies of random variable $X$, $Y$ and $H(X,Y)$ is the joint Shannon entropy. It can also be expressed in terms of the conditional entropy \cite{nielsen00}. It is to be noted that MI is always a non-negative quantity for any random variable. On the other hand, the two-point correlation function between two random variables measures the statistical overlapping between $X$ and $Y$, that is, it probes the relation between two random variables $X$ and $Y$ over a set of possible outcomes. The classical two-point correlation function is defined as \cite{ross2002first}
\begin{eqnarray}\label{eq0}
\sigma(X,Y)=\langle X Y\rangle - \langle X\rangle \langle Y\rangle
\end{eqnarray}	
where $\langle X\rangle,~ \langle Y\rangle$ are the statistical averages of $X$ and $Y$ over the probability distribution of the random variables $X$ and $Y$, and $\langle XY\rangle$ is the average of their product. The above form is often referred to as the connected or true part of the two-point correlation function as it removes the trivial contributions which arises from their statistical averages. Eq.\eqref{eq0} can lead to two possible scenarios. If $\sigma(X,Y)=0$, one can say that $X$ and $Y$ are completely independent random variables. On the other hand, for $\sigma(X,Y)\neq 0$, $X$ and $Y$ can be said to be statistically dependent. In the framework of information theory, $I(X:Y)$ captures both linear and non-linear relations, however, $\sigma(X,Y)$ is only able to capture the linear relations \cite{Knig2017QuantumEA,1990JSP}. This observation gets manifested with the help of the following bound
\begin{eqnarray}\label{Main}
I(X:Y)\geq \frac{\sigma(X,Y)^2}{2||X||^2~||Y||^2}
\end{eqnarray}
where, $||\mathcal{O}||=\sqrt{\langle \mathcal{O}^2\rangle}$ denotes the operator norm. The derivation of this inequality in context of classical information theory have been provided in Appendix \eqref{Append1}. On the other hand, the quantum mechanical version of the above inequality has the following form \cite{PhysRevLett.100.070502}
\begin{eqnarray}
I(A:B)\ge\frac{(\expval{\mathcal{O}_{A}\mathcal{O}_{B}}-\expval{\mathcal{O}_{A}}\expval{\mathcal{O}_{B}})^2}{2||\mathcal{O}_{A}||^2||\mathcal{O}_{B}||^2}
\end{eqnarray}
where $\mathcal{O}_{A}$ and $\mathcal{O}_{B}$ are some operators and $\expval{\mathcal{O}_{A}}$, $\expval{\mathcal{O}_{B}}$ are the expectation values of these operators.
Let us now have a grasp on the foundational aspect of the above inequality in the context of quantum mechanics. Consider a pure bipartite system $A\cup B$, with Hilbert space $\mathcal{H}_{A}\otimes\mathcal{H}_{B}$. So the system can be described by the density matrix $\rho_{AB}$ given by 
\begin{eqnarray}
\rho_{AB}=\ket{\psi_{AB}}\bra{\psi_{AB}}~;~\ket{\psi_{AB}}\in\mathcal{H}_{A}\otimes\mathcal{H}_{B}~.
\end{eqnarray}
Therefore, one can define the entanglement entropy of the subsystem $A$ as
\begin{eqnarray}
S_{vN}(A)=-\tr(\rho_{A}\log\rho_{A})~;~\rho_{A}=\tr_{B}(\rho_{AB})
\end{eqnarray}
where $\rho_{A}$ is the reduced density matrix of the subsystem $A$, which can be obtained by taking the trace of the total density matrix with respect to the subsystem $B$.  The quantum mutual information between two subsystems
\begin{eqnarray}
I(A:B)=S_{vN}(A)+S_{vN}(B)-S_{vN}(AB)
\end{eqnarray}
where 
\begin{eqnarray}
S_{vN}(B)=-\tr(\rho_{B}\log\rho_{B});~S_{vN}(AB)=-\tr(\rho_{AB}\log\rho_{AB})\nonumber\\
\end{eqnarray}
are von-Neumann entropies of the subsystem $B$ and the composite system $AB\equiv A\cup B$. The connected correlation between two observables $\mathcal{O}_{A}\in \mathcal{H}_A$  and $\mathcal{O}_{B}\in \mathcal{H}_B$ is given by
\begin{eqnarray}
\mathcal{C}(\mathcal{O}_{A},\mathcal{O}_{B})&=&\expval{\mathcal{O}_{A}\otimes\mathcal{O}_{B}}-\expval{\mathcal{O}_{A}}\expval{\mathcal{O}_{B}}\nonumber\\
&=&\tr(\rho_{AB}~\mathcal{O}_{A}\otimes\mathcal{O}_{B})-\tr(\rho_{A}\mathcal{O}_{A})\tr(\rho_{B}\mathcal{O}_{B})~.\nonumber\\
\end{eqnarray}
The quantum mutual information and connected correlation function can be related by expressing the mutual information in terms of the relative entropy 
\begin{eqnarray}\label{MIdef}
S(\rho_{AB}|\rho_{A}\otimes\rho_{B})=I(A:B)~.
\end{eqnarray}
Further, one has to make use of the Pinsker's inequalities \cite{ohya1993quantum,wilde2017quantum}
\begin{eqnarray}
S(\rho|\sigma)\ge\frac{1}{2}||\rho-\sigma||_1^2;~||\rho-\sigma||_1\ge\frac{\mathrm{Tr}[\mathcal{O}(\rho-\sigma)]}{||\mathcal{O}||}
\end{eqnarray}
where $||\mathcal{O}||_1$ is the trace norm of the operator $\mathcal{O}$ and $||\mathcal{O}||$ is the Hilbert-Schimdt norm of some operator $\mathcal{O}$ with the definition $\sqrt{\Tr(\mathcal{O}^{\dagger}\mathcal{O})}$ and $\Tr(\mathcal{O}(\rho-\sigma))$ is the Frobenius inner product between $(\rho-\sigma)$ and $\mathcal{O}$. If we now identify $\rho=\rho_{AB}$ and $\sigma=\rho_{A}\otimes\rho_{B}$ and $\mathcal{O}=\mathcal{O}_A\otimes\mathcal{O}_B$ in the Pinsker's inequalities, then one can write the following
 \begin{eqnarray}
 ||\rho_{AB}-\rho_{A}\otimes\rho_{B}||_1&\geq&\frac{|\expval{\mathcal{O}_{A}\mathcal{O}_B}-\expval{\mathcal{O}_A}\expval{\mathcal{O}_B}|}{||\mathcal{O}_A\otimes\mathcal{O}_B||}\nonumber\\
 &=& \frac{|\expval{\mathcal{O}_{A}\mathcal{O}_B}-\expval{\mathcal{O}_A}\expval{\mathcal{O}_B}|}{||\mathcal{O}_A||~||\mathcal{O}_B||}~.
 \end{eqnarray}
We now take square on both sides and use the definition of mutual information (given in eq.(\ref{MIdef})) to obtain the following final form \cite{susskind2014quantum}
\begin{eqnarray}\label{qid}
I(A:B)\ge\frac{(\expval{\mathcal{O}_{A}\mathcal{O}_{B}}-\expval{\mathcal{O}_{A}}\expval{\mathcal{O}_{B}})^2}{2||\mathcal{O}_{A}||^2||\mathcal{O}_{B}||^2}~.
\end{eqnarray}
We would now like to mention that the Hilbert-Schimdt norm of the operators $\mathcal{O}_{A},~\mathcal{O}_{B}$ can be defined in the following way. Let us define the inner product on the joint Hilbert space $\mathcal{H}_{AB}$ as
\begin{eqnarray}\label{innerProd}
(\mathcal{O}_{A},\mathcal{O}_{B})\equiv\mathrm{Tr}(\rho_{AB}\mathcal{O}_{A}^\dagger\mathcal{O}_{B})~.
\end{eqnarray}
Now we can define the norm of an operator as 
\begin{eqnarray}
||\mathcal{O}_{A}||\equiv \sqrt{(\mathcal{O}_{A},\mathcal{O}_{A})}=\sqrt{\Tr(\rho_{A}\mathcal{O}_{A}^\dagger\mathcal{O}_{A})}=\sqrt{\expval{\mathcal{O}_{A}^2}}\nonumber\\
\end{eqnarray}
where it has been assumed that the operator $\mathcal{O}_{A}$ is a self-adjoint operator.
Keeping the above discussion in mind, we can also rewrite the inequality (given in eq.(\ref{qid})) as
\begin{eqnarray}\label{Main2}
I(A:B)\ge\frac{(\expval{\mathcal{O}_{A}\mathcal{O}_{B}}-\expval{\mathcal{O}_{A}}\expval{\mathcal{O}_{B}})^2}{2\expval{\mathcal{O}_{A}^2}\expval{\mathcal{O}_{B}^2}}
\end{eqnarray}
where the quantity on the right hand side can be interpreted as the normalized connected two-point correlator. The expressions given in eq.(s)\eqref{Main},\eqref{Main2} suggests that the mentioned inequality along with the associated quantities, that is, mutual information and correlation are well-defined in both classical and quantum regimes. Furthermore, it is to be noted that the above discussion is restricted to zero temperature.\\
Keeping in mind the above discussed origin of the inequality, one may ask the following question. What is the status of the above inequality when we move onto a quantum field theoretic description at a finite temperature? An explicit investigation of the mentioned inequality at the level of quantum field theory is yet to be reported. In this work, we will try to address this question by using the framework of gauge/gravity duality for a large-$N$ quantum field theory (QFT) which has a holographical dual gravitational description  \cite{Maldacena:1997re,Gubser:1998bc,Witten:1998qj,Aharony:1999ti,Nastase:2015wjb,Natsuume:2014sfa}.\\
As we know, the study of information theoretic quantities get very interesting once we consider the quantum many body scenario where the famous area law of von Neumann entropy appears \cite{PhysRevD.34.373,Srednicki:1993im}. This is turn means that MI, which is expressed as a combination of the von Neumann entropies, also scales as area \cite{PhysRevLett.100.070502}. In recent times, the gauge/gravity framework has led us to a very effective approach to study information theoretic quantities for large-$N$ field theories QFTs with holographic dual. In \cite{Ryu:2006bv}, the von Neumann entropy corresponding to a subsystem of a 2d conformally invariant QFT was computed. The key idea behind this so-called holographic computation lies in the fact that the von Neumann entropy of a region $``A"$ is to be computed by evaluating the area of a minimal surface $\gamma_A$ where $\partial\gamma_A=\partial A$. On the other hand, in \cite{Balasubramanian:1999zv,Banks:1998dd,PhysRevD.62.044041}, it was proposed that for a large-$N$ QFT which has a bulk description, the two-point correlator $\langle\mathcal{O}_A(t,x)\mathcal{O}_B(t,y)\rangle$ can be obtained from the following path integral in the bulk
\begin{eqnarray}\label{general2pt}
\langle\mathcal{O}(t,x)\mathcal{O}(t,y)\rangle=\int \mathcal{DP}~e^{-\Delta L(\mathcal{P})}
\end{eqnarray}
where $L(\mathcal{P})$ is the path in the bulk geometry which connects the boundary operators and $\Delta$ is the conformal dimension of the operators. Generally, in the context of AdS/CFT duality, $\Delta$ can be expressed in the following way \cite{Gubser:1998bc,Witten:1998qj}
\begin{eqnarray}
\Delta_{\pm} =\frac{d}{2}\pm\sqrt{\frac{d^2}{4}+m^2R^2}
\end{eqnarray}
where $m$ denotes the mass of the field minimally coupled to the concerned spacetime in AdS$_{d+1}$ and $d$ is the spatial dimension. However, the formula (given in eq.(\ref{general2pt})) is valid for large conformal dimension, that is, $\Delta\sim mR$. It is to be mentioned that whether the above two-point correlator is a thermal one or not depends upon the fact that whether the bulk geometry is a blackhole spacetime or not. As we have mentioned already, we do not want to stick only to zero temperature and in order to implement this we will consider finite temperature scenario which corresponds to computation of the correlator for a thermal state. Now, from the foundational perspective, inclusion of thermal state modifies the inner-product given in eq.(\ref{innerProd}) to the thermal Wightman inner-product \cite{Bellac:2011kqa,Marolf:2013ioa}
\begin{eqnarray}
	\left(\mathcal{O}_A,\mathcal{O}_B\right)_{\beta}=\frac{1}{\mathcal{Z}}\Tr(e^{-\beta H}\mathcal{O}_A^{\dagger}\mathcal{O}_B)
\end{eqnarray}
where $\beta$ is the inverse temperature and $\mathcal{Z}=\Tr(e^{-\beta H})$ is the thermal partition function. Modification of the inner-product leads to the following thermal version of the inequality given in eq.(\ref{Main2}), namely,
\begin{eqnarray}\label{ThermalMain}
I(A:B)\ge\frac{(\expval{\mathcal{O}_{A}\mathcal{O}_{B}}_{\beta}-\expval{\mathcal{O}_{A}}_{\beta}\expval{\mathcal{O}_{B}}_{\beta})^2}{2\expval{\mathcal{O}_{A}^2}_{\beta}\expval{\mathcal{O}_{B}^2}_{\beta}}~.
\end{eqnarray}
Now, for a finite temperature QFT set up we can have the following realizations. In the above form of the inequality, $\expval{\mathcal{O}_{A}\mathcal{O}_{B}}_{\beta}\equiv\expval{\mathcal{O}(t_1,x)\mathcal{O}(t_2,y)}_{\beta}$ denotes thermal two-point correlator of field operators $\mathcal{O}(t_1,x)$, $\mathcal{O}(t_2,y)$. On the other hand, $\expval{\mathcal{O}_{A}}_{\beta} \equiv \expval{\mathcal{O}(t_1,x)}_{\beta}$ and $\expval{\mathcal{O}_{B}}_{\beta}\equiv\expval{\mathcal{O}(t_2,y)}_{\beta}$ represent the thermal one-point functions of the corresponding field points. Furthermore, $\expval{\mathcal{O}_{A}^2}_{\beta}\equiv\expval{\mathcal{O}^2(t_1,x)}_{\beta}$, $\expval{\mathcal{O}_{B}^2}_{\beta}\equiv\expval{\mathcal{O}^2(t_2,y)}_{\beta}$ are thermal norms of the field operators.\\
We now specify our bulk geometry which will be required to compute the above mentioned thermal objects holographically. In this work, we will consider the Lifshitz$_{d+1}$ ($Lif_{d+1}$) black hole spacetime \cite{Balasubramanian:2008dm}. The motivation to consider the Lifshitz black hole geometry for our work lies in the fact that it holographically enables us to probe wide range of large-$N$ QFTs. For instance, for dynamical scaling exponent $\xi=1$, one recovers SAdS$_{d+1}$ spacetime which holographically corresponds to a relativistic, large-$N$ QFT with conformal symmetry at the boundary. On the other hand, for $\xi\neq1$, we have a non-relativistic but large-$N$ QFT at the boundary.\\
The organisation of this paper is as follows. In Sec.(\ref{Sec1}), we holographically compute the relevant thermal objects required to probe the mentioned inequality, by considering $Lif_{d+1}$ black hole spacetime as the bulk geometry. In Sec.(\ref{Sec2}), we then obtain explicit expressions for the normalized connected thermal two-point correlator and mutual information and check the inequality. We conclude in Sec.(\ref{Sec3}).

\section{Holographic setup for the geometric computations}\label{Sec1}
In this section we specify the explicit geometric formulae needed for the holographic computations of the relevant objects. Firstly, we would like to mention that in this work we are interested in the equal time-correlators. This in turn means that we will consider correlation between two spatially separated fields at equal time $t$, that is, $\expval{\mathcal{O}(t,x)\mathcal{O}(t,y)}_{\beta}$. Before we proceed further we write down the geometric properties of the bulk spacetime.\\
The Lifshitz black brane geometry in $d+1$-spacetime dimensions is characterized by the following metric \cite{Balasubramanian:2008dm}
\begin{eqnarray}
ds^2=\frac{1}{z^2}\left[-\frac{f(z)}{z^{2(\xi-1)}}dt^2+\frac{dz^2}{f(z)}+\sum_{i=1}^{d-1}dx_i^2\right]
\end{eqnarray}
where $f(z)=1-\left(\frac{z}{z_+}\right)^{d-1+\xi}$, $z_+$ is the length scale corresponding to the horizon.\\
The Hawking temperature for this black brane reads 
\begin{eqnarray}
T=\left(\frac{d-1+\xi}{4\pi}\right)\frac{1}{z_+^{\xi}}~.
\end{eqnarray}
It is a well-known fact that the above geometry corresponds to a finite temperature field theory (in the large-$N$ limit) associated to the following scaling symmetry
\begin{eqnarray}
t\to\lambda^{\xi}t,~ x_i\to\lambda x_i,~ z\to \lambda^{-1}z
\end{eqnarray}
where $\xi$ is the dynamical scaling exponent \cite{Kachru:2008yh,Taylor:2008tg,Kovtun:2008qy,Chen:2009ka}. Furthermore, the Lorentz symmetry in the dual field theory is also absent (except for $\xi=1$), depicting the non-relativistic nature of the theory. This type of theories (theories with Lifshitz scaling symmetery) plays a crucial role in quantum many body physics as it provides a framework to study non-relativistic scale invariance near critical points \cite{PhysRevB.72.024420,Sachdev:2011cs}. 
\subsection{Computation of the thermal two-point correlator}
In order to compute the thermal two-point correlator (by using eq.(\ref{general2pt})), we first specify the parametrization for the path in the bulk. We assume that at $t=constant$, the bulk direction $z$ is the function of boundary direction $x_1$, that is, $z=z(x_1)$ while $z$ is independent of the rest of the directions. We then consider $dx_i=0$, where $i=2,...,d-1$. Keeping this parametrization in mind, we can write down the following
\begin{eqnarray}\label{bulkfunctional}
&&\langle\mathcal{O}(t,x)\mathcal{O}(t,y)\rangle_{\beta}\nonumber\\
&=&\int \mathcal{DP}~e^{\Delta L(\mathcal{P})}\nonumber\\
&=&\int \mathcal{D}z \exp{-\Delta \int_{0}^{|x-y|}\frac{dx_1}{z(x_1)}\sqrt{1+\frac{\left(\frac{dz(x_1)}{dx_1}\right)^2}{f(z(x_1))}}}\nonumber\\
&=&\int \mathcal{D}z~ e^{-\Delta \tilde{L}[z]}
\end{eqnarray}
where
\begin{eqnarray}
\tilde{L}[z]=\int_{0}^{|x-y|}\frac{dx_1}{z(x_1)}\sqrt{1+\frac{\left(\frac{dz(x_1)}{dx_1}\right)^2}{f(z(x_1))}}
\end{eqnarray}
is the length functional of bulk direction $z$. Now, we want to find the classical (on-shell) solution of the path integral. In order to do this we first obtain the extremized value of $\tilde{L}[z]$. This reads
\begin{eqnarray}\label{onshellL}
	L(z_t)&\equiv& \tilde{L}[z]|_{on-shell}\nonumber\\
	&=&2\int_{\epsilon}^{z_t}\frac{d\mathcal{K}}{\mathcal{K}\sqrt{1-\left(\frac{\mathcal{K}}{z_t}\right)^2}\sqrt{f(\mathcal{K})}}
\end{eqnarray}
where $z_t$ is the turning point in the bulk corresponding to the extremal bulk geodesic which is related to the spatial separation $|x-y|$ of the field operators in the following form
\begin{eqnarray}\label{xy}
|x-y|=2z_t\int_0^{z_t}\frac{\mathcal{K}d\mathcal{K}}{\sqrt{1-\left(\frac{\mathcal{K}}{z_t}\right)^2}\sqrt{f(\mathcal{K})}}~.
\end{eqnarray}
Now, substituting the on-shell value of $\tilde{L}[z]$ (given in eq.(\ref{onshellL})) in the functional integral (eq.(\ref{bulkfunctional})) and make the saddle-point approximation around the classical turning point $z=z_t$, we obtain
\begin{eqnarray}\label{EQ0}
	\langle\mathcal{O}(t,x)\mathcal{O}(t,y)\rangle_{\beta}&\sim&e^{-\Delta L(z_t)}~.
\end{eqnarray}
This is the final simplified formula which we shall use to compute the concerned thermal two-point correlator. Some recent works in which the above simplified formula has been used can be found in \cite{Rodriguez-Gomez:2021pfh,Rodriguez-Gomez:2021mkk,Krishna:2021fus}. The next step is to express it in terms of the boundary field theory parameter, that is, the separation distance $|x-y|$. In terms of this, eq.(\ref{EQ0}) becomes
 \begin{eqnarray}\label{XY}
 \langle\mathcal{O}(t,x)\mathcal{O}(t,y)\rangle_{\beta}=e^{-\Delta L(|x-y|)}~.
 \end{eqnarray}
 To proceed further, we first evaluate the integrals given in eq.(\ref{onshellL}) and eq.(\ref{xy}). We first focus on the integral for $L(z_t)$ (given in eq.(\ref{onshellL})). This reads
\begin{eqnarray}
L(z_t)&\equiv& \tilde{L}[z]|_{on-shell}\nonumber\\
&=&2\int_{\epsilon}^{z_t}\frac{d\mathcal{K}}{\mathcal{K}\sqrt{1-\left(\frac{\mathcal{K}}{z_t}\right)^2}\sqrt{f(\mathcal{K})}}\nonumber\\
&=&2\log(\frac{2z_t}{\epsilon})+\sum_{n=1}^{\infty}\frac{\Gamma[n+\frac{1}{2}]}{\Gamma[n+1]}\frac{\Gamma[\frac{n(d+\xi-1)}{2}]}{\Gamma[\frac{n(d+\xi-1)}{2}+\frac{1}{2}]}\nonumber\\
&&\times\left(\frac{z_t}{z_+}\right)^{n(d+\xi-1)}~.
\end{eqnarray}
A regularized expression for $L(z_t)$ can be obtained from the above result by subtracting the UV-regulator term, that is, $L^{(R)}(z_t)=L(z_t)-2\log(\epsilon)$. This in turn yields
\begin{eqnarray}\label{length}
L^{(R)}(z_t)=2\log(2z_t)+\sum_{n=1}^{\infty}\frac{\Gamma[n+\frac{1}{2}]}{\Gamma[n+1]}&&\frac{\Gamma[\frac{n(d+\xi-1)}{2}]}{\Gamma[\frac{n(d+\xi-1)}{2}+\frac{1}{2}]}\nonumber\\
&&\times\left(\frac{z_t}{z_+}\right)^{n(d+\xi-1)}~.
\end{eqnarray}	
On the similar note, the integral given in eq.(\ref{xy}) results in the following expression
\begin{eqnarray}\label{Tpoint}
|x-y|=2z_t+z_t\sum_{n=1}^{\infty}\frac{\Gamma[n+\frac{1}{2}]}{\Gamma[n+1]}&&\frac{\Gamma[\frac{n(d+\xi-1)}{2}+1]}{\Gamma[\frac{n(d+\xi-1)}{2}+\frac{3}{2}]}\nonumber\\
&&\times\left(\frac{z_t}{z_+}\right)^{n(d+\xi-1)}~.
\end{eqnarray}
From the expression of $|x-y|$ it can be observed that generally it is not possible to express the turning point $z_t$ in terms of $|x-y|$. However, it can be done under certain limits. 
In the low temperature limit ($|x-y|T^{1/\xi}\ll1$), we only consider terms upto $\sim \left(\frac{z_t}{z_+}\right)$. Under this approximation, the relation given in eq.\eqref{Tpoint} can be inverted to obtain the turning point in terms of the spatial separation $|x-y|$ which can further be used in eq.\eqref{length} to obtain the regularized geodesic length $L^{(R)}(|x-y|)$.
\begin{widetext}
\noindent The expression for this reads
\begin{eqnarray}
L^{(R)}(|x-y|)&=&2\log(|x-y|)+2\left(1-\frac{\sqrt{\pi}}{2^{d+1+\xi}}\frac{\Gamma[\frac{d+\xi-1}{2}+1]}{\Gamma[\frac{d-1+\xi}{2}+\frac{3}{2}]}\left[\frac{1}{d+\xi}\right]\left(\frac{|x-y|}{z_+}\right)^{(d-1)+\xi}\right)\nonumber\\
&+&\left(\frac{\sqrt{\pi}}{2^{d+\xi}}\right)\frac{\Gamma[\frac{d+\xi-1}{2}]}{\Gamma[\frac{d-1+\xi}{2}+\frac{1}{2}]}\left(\frac{|x-y|}{z_+}\right)^{(d-1)+\xi}~.
\end{eqnarray}
Finally, the desired result is to be obtained by substituting the above in the formula (eq.(\ref{XY})). This reads
\begin{eqnarray}\label{lowT2pt}
\langle\mathcal{O}(t,x)\mathcal{O}(t,y)\rangle_{\beta}=\frac{1}{|x-y|^{2\Delta}}[1-\Delta a_d \{|x-y| T^{\frac{1}{\xi}}\}^{d+\xi-1}]~.
\end{eqnarray}	
\end{widetext}	
On the other hand, in the high temperature limit ($|x-y|T^{1/\xi}\gg1$), incorporating the method shown in \cite{Fischler:2012ca}, we obtain the following result
\begin{eqnarray}\label{highT2pt}
\langle\mathcal{O}(t,x)\mathcal{O}(t,y)\rangle_{\beta}=b_dT^{\frac{2\Delta}{\xi}}e^{-\frac{|x-y|}{\chi_c}}
\end{eqnarray}
where $\chi_c=\left[\frac{d+\xi-1}{4\pi}\right]^{1/\xi}\frac{1}{\Delta T^{1/\xi}}$ denotes the thermal correlation length. The expressions corresponding to the constants $a_d$ and $b_d$ are given in the Appendix (\ref{Sec6}). Similar kind of approaches previously have been used in \cite{Keranen:2016ija,Park:2022mxj}. The above obtained results of the thermal two-point correlators are the ones obtained other under the geodesic approximation (valid for $\Delta\sim mR$). We will use these results later in order to investigate the mentioned inequality (eq.(\ref{ThermalMain})).
\subsection{Computation of the thermal one-point function}
Recently in \cite{Grinberg:2020fdj}, it has been proposed that the thermal one-point function of the CFT at the boundary can be computed from geodesic approximation. This reads
\begin{eqnarray}\label{1pt}
\langle\mathcal{O}\rangle_{\beta}\sim e^{-\Delta (l_h+i\tau_s)}~.
\end{eqnarray}
In the above expression, $l_h$ denotes the regularized proper length from boundary to the horizon and $\tau_s$ represents the proper time taken to reach the singularity from the horizon. Some recent interesting studies in this direction can be found in \cite{Berenstein:2022nlj,David:2022nfn}. In this work, we intend to compute the bound given in eq.\eqref{Main} in the gauge/gravity set up, making use of holographic techniques to compute the required quantities. Furthermore, we are also interested to see what new observations can be made by investigating this bound on information (or in correlation) as it has been never been computed explicitly in the holographic set up.
We now move on to follow the geodesic approximation approach shown in \cite{Grinberg:2020fdj} to compute the thermal one-point function (given in eq.\eqref{1pt}). The regularized proper length $l_h$ reads
\begin{eqnarray}
l_h&=&\lim_{\epsilon\rightarrow 0}\int_{\epsilon}^{z_+}\frac{dz}{z\sqrt{1-\left(\frac{z}{z_+}\right)^{d+\xi-1}}}\nonumber\\
&=&\log z_++\left(\frac{1}{d-1+\xi}\right)\log 4~.
\end{eqnarray}
On the other hand, proper time to singularity reads
\begin{eqnarray}
\tau_s=\int_{z_+}^{\infty}\frac{dz}{z\sqrt{\left(\frac{z}{z_+}\right)^{d-1+\xi}-1}}=\frac{\pi}{d-1+\xi}~.
\end{eqnarray}
Now by substituting $l_h$ and $\tau_s$ in eq.\eqref{1pt}, we obtain the following result
\begin{eqnarray}\label{1ptLif}
\langle\mathcal{O}\rangle_{\beta}=2^{2\Delta (\frac{1}{\xi}-\frac{1}{d+\xi-1})}\bigg[\frac{\pi T}{d+\xi-1}\bigg]^{\frac{\Delta}{\xi}}\times e^{-i\frac{\pi\Delta}{d-1+\xi}}~.
\end{eqnarray}
We will use the above expression in the subsequent analysis.
\subsection{Computation of $\langle\mathcal{O}^2\rangle_{\beta}$}
One can follow two different approaches in order to holographically compute the thermal object $\langle\mathcal{O}^2\rangle_{\beta}$. One of the approaches relies upon the computation of the thermal one-point function discussed above. In \cite{Grinberg:2020fdj,David:2022nfn}, it was shown that the bilinear, composite operator $\langle\mathcal{O}^2\rangle$ can be written down (under the geodesic approximation associated to a heavy gauge invariant operator) as
\begin{eqnarray}
\langle\mathcal{O}^2\rangle_{\beta}=\langle\mathcal{O}\rangle_{\beta}^2\sim e^{-2\Delta (l_h+i\tau_s)}~.	
\end{eqnarray}
The above form suggests that one can write down the following
\begin{eqnarray}\label{compositeLif}
\langle\mathcal{O}^2\rangle_{\beta}=\langle\mathcal{O}\rangle_{\beta}^2= 2^{4\Delta (\frac{1}{\xi}-\frac{1}{d+\xi-1})}\bigg[\frac{\pi T}{d+\xi-1}\bigg]^{\frac{2\Delta}{\xi}}
\end{eqnarray}
where we have ignored the overall phase factor.\\
On the other hand, one can proceed to compute  $\langle\mathcal{O}^2\rangle_{\beta}$ from the geodesic approximation formula for thermal two-point correlator. It can be observed that if we consider that both of the field operators in the thermal two-point correlator are at the same spatial point (say $x^{\prime}$), we get $\langle\mathcal{O}(t,x^{\prime})\mathcal{O}(t,x^{\prime})\rangle_{\beta}\equiv\langle\mathcal{O}^2\rangle_{\beta}$. Incorporating this argument in the low temperature expression (given in eq.(\ref{lowT2pt})), we get
 \begin{eqnarray}\label{eq36}
 &&\langle\mathcal{O}(t,x)\mathcal{O}(t,y)\rangle_{\beta}\vert_{x=y}\nonumber\\
 &=& \frac{1}{|x-y|^{2\Delta}}-\Delta a_d \left(|x-y|^{d+\xi-1-2\Delta}\right)T^{\frac{d+\xi-1}{\xi}}\vert_{x=y}~.\nonumber\\
 \end{eqnarray}  
The UV divergence coming from the first term can be removed by regulating the expression in the following way
\begin{eqnarray}
	&&\langle\mathcal{O}(t,x)\mathcal{O}(t,y)\rangle^{(R)}_{\beta}\vert_{x=y}\nonumber\\
	&:=&|\langle\mathcal{O}(t,x)\mathcal{O}(t,y)\rangle_{\beta}-\langle\mathcal{O}(t,x)\mathcal{O}(t,y)\rangle_{\beta=\infty}\vert_{x=y}\nonumber\\
	&=&\Delta a_d \left(|x-y|^{d+\xi-1-2\Delta}\right)T^{\frac{d+\xi-1}{\xi}}\vert_{x=y}~.
\end{eqnarray}
where in the second line we have subtracted the zero temperature piece from the result given in eq.(\ref{eq36}). Furthermore, if we now choose the following condition between the conformal dimension $\Delta$, spatial dimension $d$ and dynamical exponent $\xi$
\begin{eqnarray}\label{relation}
2\Delta=d+\xi-1
\end{eqnarray}
we observe something interesting. Firstly, the resultant expression is free from spatial separation $|x-y|$ and it also has the same temperature scaling we have obtained in eq.(\ref{compositeLif}), that is, $\sim T^{\frac{2\Delta}{\xi}}$. The resultant expression is 
\begin{eqnarray}\label{normT1}
\langle\mathcal{O}^2\rangle_{\beta}=\Delta a_d T^{\frac{2\Delta}{\xi}}.
\end{eqnarray}
It is to be rememebered that the above expression is only valid as long as the relation given in eq.(\ref{relation}) holds.\\
Furthermore, we can also proceed to compute $\langle\mathcal{O}^2\rangle_{\beta}$ from the high temperature expression given in eq.(\ref{highT2pt}). In this case it is pretty straight forward to show the following result
\begin{eqnarray}\label{normT2}
\langle\mathcal{O}^2\rangle_{\beta}=b_dT^{\frac{2\Delta}{\xi}}
\end{eqnarray}
where we have just used $x=y$ in eq.(\ref{highT2pt}).\\
We now have all quantities required to compute the explicit expression for the quantity
\begin{eqnarray}\label{ThermalC}
\frac{\left[\langle\mathcal{O}(t,x)\mathcal{O}(t,y)\rangle_{\beta}-\langle\mathcal{O}\rangle_{\beta}^2\right]^2}{2\langle\mathcal{O}^2\rangle^2_{\beta}}\equiv C.
\end{eqnarray}
We now focus on the computation of the MI by incorporating the results obtained from holography, and proceed to check the bound given in eq.\eqref{ThermalMain}. Before we do that we sum up all the results of $\langle\mathcal{O}^2\rangle_{\beta}$ we have obtained so far. These read
\begin{eqnarray}
 \langle\mathcal{O}^2\rangle_{\beta}&=&2^{4\Delta (\frac{1}{\xi}-\frac{1}{d+\xi-1})}\bigg[\frac{\pi T}{d+\xi-1}\bigg]^{\frac{2\Delta}{\xi}}\\
\langle\mathcal{O}^2\rangle_{\beta}&=&b_dT^{\frac{2\Delta}{\xi}}\\
\langle\mathcal{O}^2\rangle_{\beta}&=&\Delta a_d T^{\frac{2\Delta}{\xi}},~\mathrm{for}~2\Delta=d+\xi-1~.
\end{eqnarray}
This in turn means that we have three different choices for the thermal norm of the operators. We shall use all of them in the subsequent analysis and see what novel insight they lead us to.
\subsection{Computation of mutual information $I(A:B)$}
We consider two subsystems namely, $A$ and $B$ where both of these subsystems have length $l$ and widths on the other directions are fixed as $L$ \cite{PhysRevD.87.126012}. This is a standard set up for holographic computation of MI. Furthermore, we assume that $A$ and $B$ are spatially separated by a distance $s$. Keeping in mind this set up, we proceed to compute the following familiar form of MI 
\begin{eqnarray}
I(A:B)&=&S_{vN}(A) + S_{vN}(B)- S_{vN}(A\cup B)\nonumber\\
&=&2S_{vN}(l)-S_{vN}(2l+s)-S_{vN}(s)~.\nonumber\\
\end{eqnarray}
In the above expression, $S_{vN}(.)$ denote the von Neumann entropies associated to the concerned subsystems which are to be computed by using the Ryu-Takayanagi formula \cite{Ryu:2006bv,Ryu:2006ef}. Similar to the computation of the thermal two-point correlator, we holographically compute the expression of MI in both low and high temperature limits. Some previous studies related to the holographic computation of MI for Lifshitz type of geometries can be found in \cite{BabaeiVelni:2019pkw,Saha:2021kwq}. In the low temperature limit ($sT^{1/\xi}\ll lT^{1/\xi}\ll1$), we obtain the following result
\begin{widetext}
	\begin{eqnarray}\label{lowTMI}
	I(A:B)=I(A:B)|_{T=0}+\frac{c_2}{4}\left(LT^{1/\xi}\right)^{(d-2)}\left[2-\alpha^{1+\xi}-(2+\alpha)^{1+\xi}\right]\left(lT^{1/\xi}\right)^{1+\xi};~\alpha\equiv\frac{s}{l}
	\end{eqnarray}
where $I(A:B)|_{T=0}$ is the temperature independent piece which has the following expression
\begin{eqnarray}
I(A:B)|_{T=0}=\left(\frac{c_1}{4}\right)\left(\frac{LT^{1/\xi}}{lT^{1/\xi}}\right)^{d-2}\times\left[2-\frac{1}{\alpha^{d-2}}-\frac{1}{(2+\alpha)^{d-2}}\right]~.
\end{eqnarray}
We have provided the detailed expressions corresponding to the constants $c_1$ and $c_2$ in the Appendix.
On the other hand, in the high temperature limit ($sT^{1/\xi}\ll1\ll lT^{1/\xi}$), the result for MI is obtained to be
\begin{eqnarray}\label{highTMI}
I(A:B)= \frac{\left(LT^{1/\xi}\right)^{d-2}}{4}\left[c_3-\frac{c_1}{(sT^{1/\xi})^{d-2}}-\left(\frac{4\pi}{d-1+\xi}\right)^{\frac{d-1}{\xi}}(sT^{1/\xi})-c_2(sT^{1/\xi})^{1+\xi}\right]~.
\end{eqnarray}
\end{widetext}
The expression associated to the constant $c_3$ is given in the Appendix (\ref{Sec6}). We now have all the necessary expressions required for our next analysis which would be a graphical approach. We first choose a particular scenario (low temperature or high temperature) and plot the two independent quantities $C$ and see what happens.
\section{Checking the inequality and results}\label{Sec2}
In order to check the inequality given in eq.(\ref{ThermalMain}), we first consider that the spatial separation between the field operators is identified as $|x-y|=2l+s$. We shall use this for both low and high temperature expressions given in eq.(\ref{lowT2pt}) and eq.(\ref{highT2pt}). Keeping this consideration in mind, we first explicitly write down the explicit expressions for a case by case basis due to various different forms for the thermal object $\langle\mathcal{O}^2\rangle_{\beta}$ we have obtained. This reads\\

\subsection{Case 1: For $\langle\mathcal{O}^2\rangle_{\beta}=2^{4\Delta (\frac{1}{\xi}-\frac{1}{d+\xi-1})}\bigg[\frac{\pi T}{d+\xi-1}\bigg]^{\frac{2\Delta}{\xi}}$}
In this case, we consider that the form of the thermal object $\langle\mathcal{O}^2\rangle_{\beta}$ obtained in eq.(\ref{compositeLif}). Keeping this in mind, one can now proceed to obtain the explicit expression for $C$ at both low and high temperature. First, we choose the low temperature scenario where the necessary expression for the two-point correlation function is given in eq.\eqref{lowT2pt} and MI is given in eq.\eqref{lowTMI}, whereas the expressions for $\langle\mathcal{O}\rangle$ and $\langle\mathcal{O}^2\rangle$ (given in eq.(s)\eqref{1ptLif},\eqref{compositeLif}) are valid for both low and high temperature scenarios. As we are interested in comparison between the two mentioned quantities we set $|x-y|=2l(1+\alpha)$ in eq.\eqref{lowT2pt}. This leads to the following form of the two-point correlator
 \begin{widetext}
 	\begin{eqnarray}
 		\langle\mathcal{O}(t,0)\mathcal{O}(t,2l+s)\rangle_{\beta}=\frac{1}{l^{2\Delta}(2+\alpha)^{2\Delta}}\times[1-\Delta a_d(lT^{\frac{1}{\xi}})^{d+\xi-1} (2+\alpha)^{d+\xi-1}]~.	
 	\end{eqnarray}
 	We now make use of this expression along with the ones given in eq.(s)\eqref{1ptLif},\eqref{compositeLif} to obtain the following
 	\begin{eqnarray}
 		C&=&\frac{\left[\langle\mathcal{O}(t,0)\mathcal{O}(t,2l+s)\rangle_{\beta}-\langle\mathcal{O}\rangle_{\beta}^2\right]^2}{2\langle\mathcal{O}^2\rangle_{\beta}^2}\nonumber\\
 		&=&\frac{\left[1-\Delta a_d(lT^{\frac{1}{\xi}})^{d+\xi-1} (2+\alpha)^{d+\xi-1}-2^{4\Delta (\frac{1}{\xi}-\frac{1}{d+\xi-1})}\bigg[\frac{\pi}{d+\xi-1}\bigg]^{\frac{2\Delta}{\xi}}\left(lT^{\frac{1}{\xi}}\right)^{2\Delta}\left(2+\alpha\right)^{2\Delta}\right]^2}{2^{1+8\Delta (\frac{1}{\xi}-\frac{1}{d+\xi-1})}\left(2+\alpha\right)^{4\Delta}\bigg[\frac{\pi }{d+\xi-1}\bigg]^{\frac{4\Delta}{\xi}}\left(lT^{\frac{1}{\xi}}\right)^{4\Delta}}~.
 	\end{eqnarray}
Next, we consider the high temperature scenario and make use of the expressions obtained in eq.\eqref{highT2pt} to write down the following
\begin{eqnarray}
C=\frac{\left[\langle\mathcal{O}(t,0)\mathcal{O}(t,2l+s)\rangle_{\beta}-\langle\mathcal{O}\rangle_{\beta}^2\right]^2}{2\langle\mathcal{O}^2\rangle_{\beta}^2}=\frac{\left[b_d\exp{-\left(\frac{4\Delta\pi}{d+\xi-1}\right)(2+\alpha)(lT)^{1/\xi}}-2^{4\Delta (\frac{1}{\xi}-\frac{1}{d+\xi-1})}\bigg[\frac{\pi}{d+\xi-1}\bigg]^{\frac{2\Delta}{\xi}}\right]^2}{2^{1+8\Delta (\frac{1}{\xi}-\frac{1}{d+\xi-1})}\bigg[\frac{\pi}{d+\xi-1}\bigg]^{\frac{4\Delta}{\xi}}}~.
\end{eqnarray}
	~\\
	\begin{figure}[!htb]
		\centering
		\begin{minipage}[t]{0.48\textwidth}
			\centering
			\includegraphics[width=\textwidth]{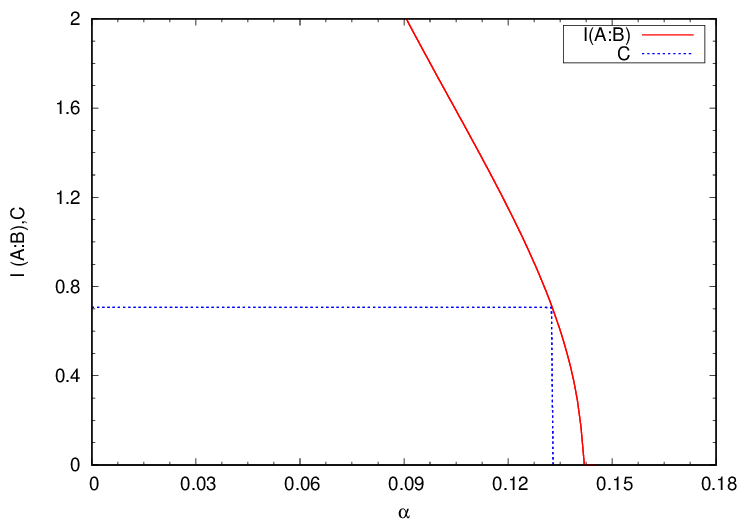}\\
			{\footnotesize  Low T scenario with $lT=0.4$.}
		\end{minipage}\hfill
		\begin{minipage}[t]{0.48\textwidth}
			\centering
			\includegraphics[width=\textwidth]{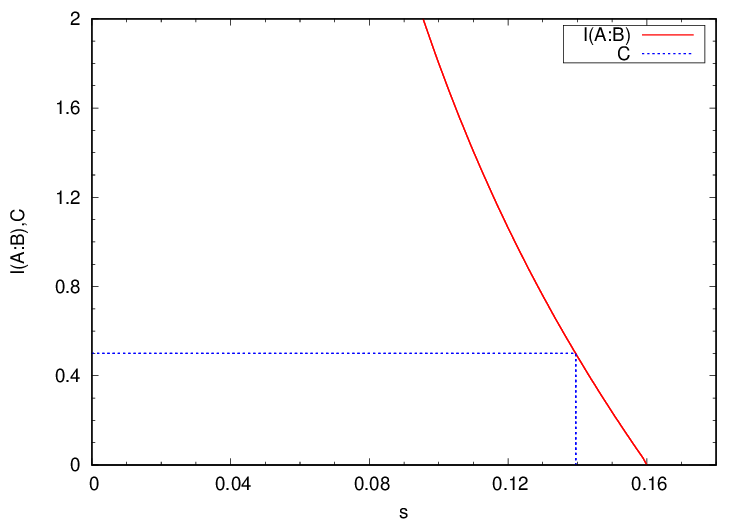}\\
			{\footnotesize  High T scenario with $lT=10$}
		\end{minipage}
		\caption{Plot of holographic mutual information $I(A:B)$ and two-point connected correlator $C$. In the left panel, we plot $I(A:B)$ and $C$ with respect to $\alpha=\frac{sT}{lT}$, for $lT=0.4$. In the right panel, we plot $I(A:B)$ and $C$ with respect to $sT$, for $lT=10$. In these plots, we set $d=3$ and $\Delta=10$.}
		\label{Plots}
	\end{figure}
\end{widetext}
We now proceed to plot the obtained expressions for the normalized connected correlator $C$ along with the ones obtained for MI (given in eq.\eqref{lowTMI} and eq.\eqref{highTMI}) for a particular value of $lT^{\frac{1}{\xi}}$ in $d=3$ dimensions. Before we proceed to the plots, we set $\xi=1$ which fixes the AdS-Schwarzschild$_{d+1}$ black brane as the bulk geometry and the dual field theory is a conformal field theory at a finite temperature. In the left panel of Fig.\eqref{Plots}, we give the plots corresponding to the low temperature scenario whereas the plots for the high temperature scenario is given in the right panel. In these plots, we observe that both $I(A:B)$ and $C$ decreases with the increase in the value of the separation distance $sT$ (in dimensionless form). Furthermore, we observe that initially $I(A:B)$ is much greater than $C$ and at a particular value of the separation distance $sT$, the quantum mutual information $I(A:B)$ is equal to the normalized connected two-point correlator $C$. We denote this specific value of the separation distance as $sT|_{c}$. However, it is to be mentioned that beyond $sT|_{c}$, $I(A:B)$ gets smaller than $C$ and the inequality gets violated. In order to avoid this situation, we make a proposal now.\\

\textit{All of the quantum mechanical dependencies (correlations) between the subsystems $A$ and $B$ vanishes once the bound (given in eq.\eqref{Main2}) gets saturated and only the classical dependencies survive which is represented by a non-zero mutual information $I(A:B)\neq0$.}\\

We shall now justify the reasoning behind the above proposal. It is a well-known fact that the mutual information quantifies not only quantum mechanical dependencies (correlations) between the subsystems but also the classical ones too. In short, none of the existing dependencies are overlooked. However, the normalized connected two-point correlator is only able to quantify the quantum mechanical dependencies between the concerned operators which belong to the associated subsystems, not the classical ones. Due to this reason it is not possible to have a correlated system with vanishing mutual information. We propose that at $sT=sT|_{c}$ the bound $I(A:B)\geq C$ gets saturated and all of the existing quantum mechanical dependencies between the operators $\mathcal{O}_A\in\mathcal{H}_A$ and $\mathcal{O}_B\in\mathcal{H}_B$ vanishes. This is manifested by the fact that at $sT=sT|_{c}$, $\frac{\left[\langle\mathcal{O}(t,0)\mathcal{O}(t,2l+s)\rangle-\langle\mathcal{O}\rangle^2\right]^2}{2\langle\mathcal{O}^2\rangle^2}=0$. However, $I(A:B)\neq0$, which in turn means that only the classical dependencies between the subsystems survive. Furthermore, these classical dependencies also vanish for another value of the separation distance, namely, $sT|_{I}$ at which the mutual information between the subsystems vanishes, that is, $I(A:B)=0$ and the subsystems become completely independent of each other. It is to be noted that $sT|_{I}>sT|_{C}$.
As we have mentioned earlier, in Fig.\eqref{Plots} we have graphically represented our observation for a AdS$_3$-Schwarzschild black brane. In the low temperature scenario we get $sT|_c\approx0.053$ and $sT|_I\approx 0.0608$, whereas in the high temperature scenario, we have $sT|_c\approx0.139$ and $sT|_I\approx 0.16$. In order to have a general observation, we now proceed to repeat the above observation for various values of the dynamical exponent $\xi$, that is, we consider $\xi=2,3,..~.$ We give the obtained values of $sT|_c$ and $sT|_I$ in a Table.\begin{table}[H]\label{Table1}
	\begin{tabular}{l|SS||SS}
		\hline
		\toprule
		\multirow{2}{*}{$\xi$} &
		\multicolumn{2}{c}{low T} &
		\multicolumn{2}{c}{high T}\\
		& {$sT|_c$} & {$sT|_I$} & {$sT|_c$} & {$sT|_I$} \\
		\hline
		\midrule
		$2$ & 0.136 & 0.160 & 0.265 & 0.372 \\
		$3$ & 0.188 & 0.201 & 0.309 & 0.475 \\
		\hline
		\bottomrule
	\end{tabular}
\end{table}
The above study suggests that in context of holographic computation of mutual information and two-point correlator, it is possible to point out two different critical separation distances which denote the vanishing of quantum mechanical dependencies and classical dependencies. It is evident that for both low and high temperature regimes, $sT|_I>sT|_c$. In the next subsections, we shall use the results of $\langle\mathcal{O}^2\rangle_{\beta}$ obtained from the thermal Wightman function (two-point correlator).

\subsection{Case 2: For $\langle\mathcal{O}^2\rangle_{\beta}=b_dT^{\frac{2\Delta}{\xi}}$}
We now once again compute the relevant expressions by considering the result $\langle\mathcal{O}^2\rangle_{\beta}=b_d T^{\frac{2\Delta}{\xi}}$ (given in eq.(\ref{normT2})).
\begin{widetext}
In this case, the low temperature expression reads
\begin{eqnarray}
C&=&\frac{\left[\langle\mathcal{O}(t,0)\mathcal{O}(t,2l+s)\rangle_{\beta}-\langle\mathcal{O}\rangle_{\beta}^2\right]^2}{2\langle\mathcal{O}^2\rangle_{\beta}^2}\nonumber\\
&=&\frac{\left[1-\Delta a_d(lT^{\frac{1}{\xi}})^{d+\xi-1} (2+\alpha)^{d+\xi-1}-2^{4\Delta (\frac{1}{\xi}-\frac{1}{d+\xi-1})}\bigg[\frac{\pi}{d+\xi-1}\bigg]^{\frac{2\Delta}{\xi}}\left(lT^{\frac{1}{\xi}}\right)^{2\Delta}\left(2+\alpha\right)^{2\Delta}\right]^2}{2b_d^2\left(2+\alpha\right)^{4\Delta} \left(lT^{\frac{1}{\xi}}\right)^{4\Delta}}
\end{eqnarray}
where we have used the expressions given in eq.(\ref{lowT2pt}), eq.(\ref{normT2}) and eq.(\ref{1ptLif}).\\ On the other hand, the high temperature expression reads
\begin{eqnarray}
C&=&\frac{\left[\langle\mathcal{O}(t,0)\mathcal{O}(t,2l+s)\rangle_{\beta}-\langle\mathcal{O}\rangle_{\beta}^2\right]^2}{2\langle\mathcal{O}^2\rangle_{\beta}^2}=\frac{\left[b_d\exp{-\left(\frac{4\Delta\pi}{d+\xi-1}\right)(2+\alpha)(lT)^{1/\xi}}-2^{4\Delta (\frac{1}{\xi}-\frac{1}{d+\xi-1})}\bigg[\frac{\pi}{d+\xi-1}\bigg]^{\frac{2\Delta}{\xi}}\right]^2}{2b_d^2}~.
\end{eqnarray}	
\end{widetext}
In computing the above, we have used the expressions given in eq.(\ref{highT2pt}), eq.(\ref{normT2}) and eq.(\ref{1ptLif}). We observe that in both the cases, $C=\frac{\left[\langle\mathcal{O}(t,0)\mathcal{O}(t,2l+s)\rangle_{\beta}-\langle\mathcal{O}\rangle_{\beta}^2\right]^2}{2\langle\mathcal{O}^2\rangle_{\beta}^2}\approx 0$. This in turn means that the all of the dependencies (quantum or classical) measured by MI vanishes altogether when $I(A:B)=C\approx 0$. If we now compare this scenario with the one we have explained in Case 1, we observe that for $\langle\mathcal{O}^2\rangle_{\beta}=b_d T^{\frac{2\Delta}{\xi}}$, both of the critical separation lengths are same which means that MI trivially vanishes and becomes equal to $C=\frac{\left[\langle\mathcal{O}(t,0)\mathcal{O}(t,2l+s)\rangle_{\beta}-\langle\mathcal{O}\rangle_{\beta}^2\right]^2}{2\langle\mathcal{O}^2\rangle_{\beta}^2}$.
\subsection{Case 3: For $\langle\mathcal{O}^2\rangle_{\beta}=\Delta a_d T^{\frac{2\Delta}{\xi}}$}
In this case, we consider the form of the thermal object $\langle\mathcal{O}^2\rangle_{\beta}$ we have obtained in eq.(\ref{normT1}) and proceed to investigate the mentioned inequality. Let us first compute the forms of the normalized connected phase of the thermal correlator $C$ in both low and high temperatures. 
\begin{widetext}
	In the low temperature limit, it now reads
	\begin{eqnarray}
	C&=&\frac{\left[\langle\mathcal{O}(t,0)\mathcal{O}(t,2l+s)\rangle_{\beta}-\langle\mathcal{O}\rangle_{\beta}^2\right]^2}{2\langle\mathcal{O}^2\rangle_{\beta}^2}\nonumber\\
	&=&\frac{\left[1-\Delta a_d(lT^{\frac{1}{\xi}})^{d+\xi-1} (2+\alpha)^{d+\xi-1}-2^{4\Delta (\frac{1}{\xi}-\frac{1}{d+\xi-1})}\bigg[\frac{\pi}{d+\xi-1}\bigg]^{\frac{2\Delta}{\xi}}\left(lT^{\frac{1}{\xi}}\right)^{2\Delta}\left(2+\alpha\right)^{2\Delta}\right]^2}{2\left(\Delta a_d\right)^2\left(2+\alpha\right)^{4\Delta} \left(lT^{\frac{1}{\xi}}\right)^{4\Delta}}
	\end{eqnarray}
	where we have used the expressions given in eq.(\ref{lowT2pt}), eq.(\ref{normT1}) and eq.(\ref{1ptLif}). On the other hand, in the high temperature limit, we now have
	\begin{eqnarray}
	C&=&\frac{\left[\langle\mathcal{O}(t,0)\mathcal{O}(t,2l+s)\rangle_{\beta}-\langle\mathcal{O}\rangle_{\beta}^2\right]^2}{2\langle\mathcal{O}^2\rangle_{\beta}^2}=\frac{\left[b_d\exp{-\left(\frac{4\Delta\pi}{d+\xi-1}\right)(2+\alpha)(lT)^{1/\xi}}-2^{4\Delta (\frac{1}{\xi}-\frac{1}{d+\xi-1})}\bigg[\frac{\pi}{d+\xi-1}\bigg]^{\frac{2\Delta}{\xi}}\right]^2}{2\left(\Delta a_d\right)^2}
	\end{eqnarray}
	where we have used the results given in eq.(\ref{highT2pt}), eq.(\ref{normT1}) and eq.(\ref{1ptLif}). Furthermore, in both of the above expressions one has to use $2\Delta=d+\xi-1$.
\end{widetext}
However, we observe that in this case or more precisely for $\langle\mathcal{O}^2\rangle_{\beta}=\Delta a_d T^{\frac{2\Delta}{\xi}}$, the inequality $I(A:B)\geq \frac{\left[\langle\mathcal{O}(t,x)\mathcal{O}(t,y)\rangle_{\beta}-\langle\mathcal{O}\rangle_{\beta}^2\right]^2}{2\langle\mathcal{O}^2\rangle^2_{\beta}}$ does not hold. The possible reason behind this may lie in the fact that $\langle\mathcal{O}^2\rangle_{\beta}=\Delta a_d T^{\frac{2\Delta}{\xi}}$ has been obtained by using the low temperature expression of the thermal two-point correlator which does not capture full thermal dependencies required for the computation.
\section{Conclusion}\label{Sec3}
We now summarize our findings. In this work we have investigated the inequality (or bound on correlations from mutual information) $I(A:B)\ge\frac{(\expval{\mathcal{O}_{A}\mathcal{O}_{B}}_{\beta}-\expval{\mathcal{O}_{A}}_{\beta}\expval{\mathcal{O}_{B}}_{\beta})^2}{2\expval{\mathcal{O}_{A}^2}_{\beta}\expval{\mathcal{O}_{B}^2}_{\beta}}$ using the gauge/gravity duality. The mentioned inequality can be derived explicitly in context of classical or quantum information theory and it has been extensively studied in these contexts. However, in the context of holographic studies of information theoretic quantities, one just assumes that the inequality must hold but a proper study (which probes the applicability or consequences of the inequality) on the inequality was not addressed in the literature. This we have tried to address in this work. To proceed, we consider a Lifshitz$_{d+1}$ black hole spacetime as the bulk geometry which holographically represents a finite temperature large-$N$ QFT with or without Lorentz invariance (depending upon the choice of the dynamical exponent) at the boundary. By utilizing this setup, we first compute the normalized connected phase of the two-point correlator, that is, $C=\frac{(\expval{\mathcal{O}(t,x)\mathcal{O}(t,y)}_{\beta}-\expval{\mathcal{O}}_{\beta}^2)^2}{2\expval{\mathcal{O}_{\beta}^2}^2}$. It can be observed that it is a collection of three different thermal objects which we need to compute in order to have an explicit expressions for it. These are two-point correlator $\expval{\mathcal{O}(t,x)\mathcal{O}(t,y)}_{\beta}$, one-point function $\expval{\mathcal{O}}_{\beta}$ and $\expval{\mathcal{O}_{\beta}^2}$ which is related to the operator norm (thermal) as $||\mathcal{O}_{\beta}||^2=\expval{\mathcal{O}_{\beta}^2}$. It is to be mentioned that in this work we have considered equal-time correlation between the field operators. As we know, the conformal dimension of the operators at the boundary are related to the mass of the dual fields in the bulk and for heavy operators with large conformal dimension, that is for $\Delta\sim mR$, one can compute $\expval{\mathcal{O}(t,x)\mathcal{O}(t,y)}_{\beta}$ from the normalized geodesic length in the bulk. This lead us to two different expressions, one valid for small temperature ($lT^{\frac{1}{\xi}}\ll 1$) and the other one valid for large temperature ($lT^{\frac{1}{\xi}}\gg 1$). Next, we holographically compute the thermal one-point function $\expval{\mathcal{O}}_{\beta}$ under the similar condition (large $\Delta$). It is to be mentioned that for $\Delta\sim mR$, $\expval{\mathcal{O}}_{\beta}$ is related to two different entities, the regulazied proper length from boundary to the horizon and the proper time taken to reach singularity from horizon. The resulting expression of $\expval{\mathcal{O}}_{\beta}$ is valid for any temperature. Computation of thermal object $\expval{\mathcal{O}_{\beta}^2}$ becomes very interesting due to the fact that it can be computed in three different methods. The first method relies upon the recent observation $\expval{\mathcal{O}_{\beta}^2}=\expval{\mathcal{O}_{\beta}}^2$, that is, it can be computed from the obtained result for thermal one-point function \cite{David:2022nfn}. The second method is to compute it from the obtained results of the thermal two-point correlator, that is, $\expval{\mathcal{O}_{\beta}^2}=\expval{\mathcal{O}(t,x)\mathcal{O}(t,y)}_{\beta}|_{x=y}$. As we have mentioned, one can have two different expressions for the thermal two-point correlator (for low and high temperature). This in turn means in this method one also obtains two different expressions for $\expval{\mathcal{O}_{\beta}^2}$, one from the low temperature expression of the thermal two-point correlator and the other one derived from the high temperature expresion of $\expval{\mathcal{O}(t,x)\mathcal{O}(t,y)}_{\beta}$. Finally, we holographically compute the mutual information $I(A:B)$ by using the well-known Ryu-Takayangi prescription for two disjoint subsystems ($A$ and $B$) of same length $l$, separated by a distance $s$. It is to be mentioned that similar to the thermal two-point correlator, $I(A:B)$ also has two different expressions due to different behaviours in the low and high temperature regimes. Once we have all of the necessary quantities, we proceed to check the mentioned inequality. We observe that in the case where we have considered $\expval{\mathcal{O}_{\beta}^2}=\expval{\mathcal{O}_{\beta}}^2$, there are two different critical separation lengths, these can be understood in the following way. At the first critical separation length $sT|_c$ mutual information equal the normalized connected phase of two-point correlator, we propose that it signifies the fact that all of the quantum dependencies (correlations) existing between the operators $\mathcal{O}A\in\mathcal{H}_A$ and $\mathcal{O}_B\in\mathcal{H}_B$ vanishes at $sT=sT|_c$. On the the other hand, there exists another critical separation length $sT|_I$ (which is greater than $sT|_c$) at which the rest of the existing classical dependencies vanishes. Further, mutual information measures all of the existing correlations (let it be quantum or classical dependencies) between the subsystems ($A$ and $B$) however, the normalized connected phase of thermal correlator is able to measure only the quantum dependencies. Keeping this in mind, we further propose that at $sT=sT|_c$, once the bound saturates, that is, $I(A:B)=\frac{(\expval{\mathcal{O}_{A}\mathcal{O}_{B}}_{\beta}-\expval{\mathcal{O}_{A}}_{\beta}\expval{\mathcal{O}_{B}}_{\beta})^2}{2\expval{\mathcal{O}_{A}^2}_{\beta}\expval{\mathcal{O}_{B}^2}_{\beta}}$,the quantity on the right hand side jumps to a vanishing value. The scenario, drastically changes once we consider the derivation of $\expval{\mathcal{O}_{\beta}^2}$ from the thermal two-point function, that is, $\expval{\mathcal{O}_{\beta}^2}=\expval{\mathcal{O}(t,x)\mathcal{O}(t,y)}_{\beta}|_{x=y}$. We observe that the computed result is ill-defined when we derive it from the low temperature expression of $\expval{\mathcal{O}(t,x)\mathcal{O}(t,y)}_{\beta}$, this may be due to the fact that it does not capture the total necessary thermal or temperature dependency (as it has been derived from the low temperature approximated expression). However, if we derive $\expval{\mathcal{O}_{\beta}^2}$ from the high temeprature result of $\expval{\mathcal{O}(t,x)\mathcal{O}(t,y)}_{\beta}$, it is well-defined and leads to the result $\frac{(\expval{\mathcal{O}_{A}\mathcal{O}_{B}}_{\beta}-\expval{\mathcal{O}_{A}}_{\beta}\expval{\mathcal{O}_{B}}_{\beta})^2}{2\expval{\mathcal{O}_{A}^2}_{\beta}\expval{\mathcal{O}_{B}^2}_{\beta}}=0$. This in turn means in this case, both of the previously mentioned critical separation lengths becomes identical, that is, $sT|_c=sT|_I$, at which $I(A:B)=\frac{(\expval{\mathcal{O}_{A}\mathcal{O}_{B}}_{\beta}-\expval{\mathcal{O}_{A}}_{\beta}\expval{\mathcal{O}_{B}}_{\beta})^2}{2\expval{\mathcal{O}_{A}^2}_{\beta}\expval{\mathcal{O}_{B}^2}_{\beta}}=0$.
\section{Appendix 1: Classical derivation of the inequality involving mutual information and connected thermal two-point correlator, $I(X:Y)\geq \frac{\sigma(X,Y)^2}{2||X||^2~||Y||^2}$}\label{Append1}
Let us consider a classical random variable $X$, with the probablity distribution function (PDF) $f(x)$. In this setup, the Shannon (differential) entropy of the classical random variable $X$ is defined as
\begin{eqnarray}\label{eq1}
H(X)=-\int_{-\infty}^{+\infty}f(x)\ln(f(x))~dx~.
\end{eqnarray}
To proceed further, we will consider the PDF to be a Gaussian distribution, given by
\begin{eqnarray}\label{eq2}
f(x)=\frac{1}{\sqrt{2\pi\sigma^2}}\exp(-\frac{(x-\mu_{X})^2}{2\mathrm{Var}_{X}^2})
\end{eqnarray}
where $\mu_{X}$ is the mean and $\mathrm{Var}_{X}$ is the variance of the random variable $X$ with respect to the Gaussian distribution. With this Gaussian probability distribution for $X$, we can compute the differential entropy of the random variable $X$ by using the  eq.(\ref{eq1}). The Shannon entropy of the random variable $X$ with the Gaussian probability distribution (given in eq.(\ref{eq2})) is obtained to be
\begin{eqnarray}
H(X)=\frac{1}{2}\ln(2\pi e \mathrm{Var}_{X}^2)~.
\end{eqnarray}
Similarly, for another random variable $Y$, with the Gaussian probability distribution $f(y)$, the differential entropy is given by
\begin{eqnarray}
H(Y)=\frac{1}{2}\ln(2\pi e \mathrm{Var}_{Y}^2)~.
\end{eqnarray}
Now, as we have mentioned earlier in the introduction of the paper, the mutual information between two random variables $X$ and $Y$ is given by eq.\eqref{CMI}. The Shannon entropy corresponding to the joint probability distribution is given by
\begin{eqnarray}
H(XY)=\frac{1}{2}\ln((2\pi e)^2\det(\Sigma))
\end{eqnarray}
where $\Sigma$ is the covariance matrix, given by
\begin{eqnarray}
\Sigma=\begin{bmatrix}
\mathrm{Var}_{X}^2& \sigma(X,Y) \\
\sigma(X,Y) & \mathrm{Var}_{Y}^2
\end{bmatrix}
\end{eqnarray}
where $\sigma(X,Y)$ is the connected part of the two-point correlation which has the form given in eq.\eqref{eq0}. Thus, the determinant of the covariance matrix is given by
\begin{eqnarray}
\det(\Sigma)=\mathrm{Var}_{X}^2\mathrm{Var}_{Y}^2-\sigma(X,Y)^2~.
\end{eqnarray}
Now, By substituting the above result in the definition of MI, we obtain the following
\begin{eqnarray}
I(X:Y)&=&\frac{1}{2}\log\Bigg(\frac{\mathrm{Var}_{X}^2\mathrm{Var}_{Y}^2}{\mathrm{Var}_{X}^2\mathrm{Var}_{Y}^2-\sigma(X,Y)^2}\Bigg)\nonumber\\
&\ge&\frac{1}{2}\left[\frac{\sigma(X,Y)^2}{||X||^2||Y||^2}\right]~.
\end{eqnarray}
In the last line, we have used the following inequality
\begin{eqnarray}
\log(1+\lambda)\ge\frac{\lambda}{1+\lambda}~.
\end{eqnarray}
This completes the proof of the classical version of the inequality given in eq.(\ref{Main}).
\section{Appendix 2: Explicit expressions of various constants}\label{Sec6}
In this Appendix, we provide the explicit expressions corresponding to various constants that we have introduced in our analysis for the sake of simplicity.
\begin{widetext}
	\begin{eqnarray}
	a_d&=&\frac{\sqrt{\pi}}{2^{d+\xi}}\left(\frac{4\pi}{d-1+\xi}\right)^{1+\frac{d-1}{\xi}}\frac{\Gamma[\frac{d+\xi-1}{2}]}{\Gamma[\frac{d-1+\xi}{2}+\frac{1}{2}]}\left[\frac{1}{d+\xi}\right]\nonumber\\
	b_d&=&\frac{e^{2\Delta}}{2^{2\Delta}}\left[\frac{4\pi}{d-1+\xi}\right]^{\frac{2\Delta}{\xi}}\exp(\sum_{n=1}^{\infty}-\left[\frac{\Delta}{n(d-1+\xi)}\right]\frac{\Gamma[n+\frac{1}{2}]}{\Gamma[n+1]}\frac{\Gamma[1+\frac{n(d-1+\xi)}{2}]}{\Gamma[\frac{3}{2}+\frac{n(d-1+\xi)}{2}]})\nonumber\\
	c_1&=&-\frac{(2\sqrt{\pi})^{d-1}}{(2d-3)}\frac{\Gamma[\frac{2-d}{2(d-1)}]}{\Gamma[\frac{3-2d}{2(d-1)}]}\left(\frac{\Gamma[\frac{d}{2(d-1)}]}{\Gamma[\frac{1}{2(d-1)}]}\right)^{d-2}\nonumber\\
	c_2&=&\left[\frac{\Gamma[\frac{1}{2(d-1)}]}{\Gamma[\frac{d}{2(d-1)}]}\right]^{1+\xi}\left[\frac{4\pi}{d-1+\xi}\right]^{1+\frac{d-1}{\xi}}\left(\frac{1}{2\sqrt{\pi}}\right)^{\xi}\left(\frac{1}{4(d-1)}\right)\times\left[\frac{\Gamma[\frac{1+\xi}{2(d-1)}]}{\Gamma[\frac{d+\xi}{2(d-1)}]}+\frac{(d-2)}{(2d-3)}\frac{\Gamma[\frac{2-d}{2(d-1)}]}{\Gamma[\frac{3-2d}{2(d-1)}]}\frac{\Gamma[\frac{2d-1+\xi}{2(d-1)}]}{\Gamma[\frac{3(d-1)+\xi}{2(d-1)}]}\right]\nonumber\\
	c_3&=&\left(\frac{4\pi}{d-1+\xi}\right)^{\frac{d-2}{\xi}}\left[\sum_{n=1}^{\infty}\frac{1}{[1+n\xi+n(d-1)]}\frac{\Gamma[n+\frac{1}{2}]}{\Gamma[n+1]}\frac{\Gamma[\frac{1+n\xi+(n-1)(d-1)}{2(d-1)}]}{\Gamma[\frac{1+n\xi+n(d-1)}{2(d-1)}]}-\sqrt{\pi}\left(\frac{d}{d-2}\right)\frac{\Gamma[\frac{d}{2(d-1)}]}{\Gamma[\frac{1}{2(d-1)}]}\right]\nonumber~.
	\end{eqnarray}
\end{widetext} 
\section*{Acknowledgements}
\noindent AS would like to thank S.N. Bose National Centre for Basic Sciences for the financial support through its Advanced Postdoctoral Research Programme. The authors would like to thank the anonymous referee for the crucial comments which has helped us to substantially improve our manuscript.
\bibliographystyle{hephys}   
\bibliography{Reference.bib}

\end{document}